%% file: paper.tex
\documentclass[ALICE,manyauthors]{cernphprep}
\usepackage[modulo]{lineno}
\usepackage{hyperref}
\usepackage[comma,square,numbers,sort&compress]{natbib}
\usepackage{amsmath,amssymb}% for advanced math typesetting 
\usepackage{mathrsfs} % produce the curly R with \mathscr not \mathcal

\input{helpers.tex}
\begin{document}
%%%%%%%%%%%%% ptdr definitions %%%%%%%%%%%%%%%%%%%%%
%
%%%%%%%%%%%%%%%  Title page %%%%%%%%%%%%%%%%%%%%%%%%
%

\newcommand{\pPb}{\textnormal{p--Pb}}
\newcommand{\RpPb}{\ensuremath{R_\mathrm{pPb}}}
\newcommand{\PbPb}{\textnormal{Pb--Pb}}
\newcommand{\AuAu}{\textnormal{Au--Au}}
\newcommand{\pp}{\ensuremath{\mbox{p}\mbox{p}}}
\newcommand{\snn}{\ensuremath{\sqrt{s_\tn{NN}}}}
\newcommand{\kt}{\ensuremath{k_\tn{T}}}\newcommand{\kT}{\kt}
\newcommand{\pt}{\ensuremath{p_\tn{T}}}\newcommand{\pT}{\pt}
\newcommand{\ptch}{\ensuremath{p_\mathrm{T,\,ch\;jet}}}
\newcommand{\deltaptch}{\ensuremath{\delta p_\mathrm{T,\,ch}}}

 % off by redefinding CKBNOTEje
\newcommand{\CKBNOTE}[1]{{\bf CKB:  #1}} 
\renewcommand{\CKBNOTE}[1]{}  % switch off

 % off by redefinding RHNOTE
\newcommand{\RHNOTE}[1]{{\bf RH:  #1}} 
\renewcommand{\RHNOTE}[1]{}  % switch off

\begin{titlepage}
\PHyear{2015}
\PHnumber{040}                 % required, obtained from PH
\PHdate{24 February}              % required, will be obtained from PH
%
%
%%% Put your own title + short title here:
\title{Measurement of charged jet production cross sections and\\ nuclear modification in p--Pb collisions at $\mathbf{\sqrt{s_\mathrm{NN}} = 5.02}$ TeV}
\ShortTitle{Charged jets in p--Pb}   % appears on right page headers

%
%%% Do not change the nexts!
\Collaboration{ALICE Collaboration%
         \thanks{See Appendix~\ref{app:collab} for the list of collaboration
                      members}}
\ShortAuthor{ALICE Collaboration}      % appears on left page headers, do not change
\begin{abstract}
Charged jet production cross sections in $\pPb$ collisions at $\snn = 5.02 \TeV$ measured with the ALICE detector at the LHC are presented. 
Using the anti-$\kt$ algorithm, jets have been reconstructed in the central rapidity region from charged particles with resolution parameters $R = 0.2$ and $R = 0.4$. 
The reconstructed jets have been corrected for detector effects and the underlying event background.
To calculate the nuclear modification factor, $\RpPb$, of charged jets in $\pPb$ collisions, a $\pp$ reference was constructed by scaling previously measured charged jet spectra at $\sqrt{s} = 7 \TeV$.
In the transverse momentum range $20 \le \ptch \le 120 \GeV$/$c$, $\RpPb$ is found to be consistent with unity, indicating the absence of strong nuclear matter effects on jet production. 
Major modifications to the radial jet structure are probed  via the ratio of jet production cross sections reconstructed with the two different resolution parameters.
This ratio is found to be similar to the measurement in $\pp$ collisions at $\sqrt{s} = 7$ TeV and to the expectations from PYTHIA $\pp$ simulations and NLO pQCD calculations at $\snn = 5.02 \TeV$.
\end{abstract}
\end{titlepage}
\setcounter{page}{2}

% \input{main.tex}               %%%%%%%%%%% put the body of the article here

% !TEX root = paper.tex

\section{Introduction}

Jets are the observable final state of a fragmenting parton produced e.g. in scattering of partons in nuclei with a large momentum  transfer, $Q^2$.  
At sufficiently large $Q^2$, the jet production cross section is
computable since it can be factorized into the non-perturbative parton distribution
and fragmentation functions and the  cross section of partonic scatterings, which is calculable
in perturbative QCD (pQCD) \cite{Col85a}.
Jet measurements in $\pPb$ and their comparison to $\pp$ provide a tool to better
constrain effects of (cold) nuclear matter on these factors. 
In particular, they can be used to examine the role of a modification of the initial distribution
of quarks and gluons, e.g. shadowing effects and gluon
saturation \cite{McLerran:2001sr,Salgado:2011wc}, and the impact of
multiple scatterings and hadronic re-interactions in the initial and
final state \cite{Krzywicki:1979gv,Accardi:2007in}.

In central heavy-ion collisions, the production of jets and high-$\pt$
particles is strongly modified: in $\PbPb$ collisions at the LHC, the
observed hadron yields are suppressed  by up to a factor of seven  compared
to $\pp$ collisions, approaching a factor of two suppression at high $\pt$
\cite{Aamodt:2010jd,Aamodt:2011vg,CMS:2012aa}.
A similar suppression is also observed for reconstructed jets in
central $\PbPb$ \cite{Aad:2010bu,Chatrchyan:2012nia,Aad:2012vca,Abelev:2013kqa,Aad:2014bxa}.
This phenomenon, referred to as \emph{jet quenching}, has also been
observed previously in high-$\pt$ particle production in central
$\AuAu$ collisions at RHIC
\cite{Adcox:2001jp,Adler:2003qi,Ada03b,Adams:2003im,Arsene:2003yk,Back:2003qr}.
It is attributed to the creation of a quark-gluon plasma (QGP) in the
final state, where hard scattered partons radiate gluons in strong
interactions  with the medium as first predicted in \cite{Gyulassy:1990ye,Baier:1994bd}.
This results in  an radiative energy loss of the
leading parton and a modified fragmentation pattern.

Initially, $\pPb$ collisions have been seen as the testing ground for
isolated cold nuclear matter effects, without the formation of a hot and dense medium.
However, recent results on low-$\pt$ particle production and long
range correlations in $\pPb$ collisions at $\sqrt{s_\mathrm{NN}} =
5.02$ TeV \cite{CMS:2012qk,Abelev:2012ola,Aad:2013fja,Abelev:2013wsa} exhibit features of collective behavior, similar to those
found in $\PbPb$ collisions, where they are attributed to the creation of a QGP. 
At high $\pt$, results on the production of unidentified charged
particles
\cite{ALICE:2012mj,Abelev:2014dsa,Khachatryan:2015xaa,ATLAS-CONF-2014-029}
and jets \cite{ATLAS:2014cpa,Chatrchyan:2014hqa} in $\pPb$ collisions at $\sqrt{s_\mathrm{NN}} = 5.02$ TeV are
consistent with the absence of a strong final state suppression. 
The question to what extent other nuclear effects lead to an enhancement of particle production at high $\pt$ is still open, a possible enhancement in $\pPb$ collisions has been reported for single charged hadrons  \cite{Khachatryan:2015xaa}.  
The measurement of jets in $\pPb$ collisions compared to single hadrons tests the parton fragmentation beyond the leading particle with the inclusion of low-$\pt$ and large-angle fragments.

A jet is defined experimentally by the algorithm that combines the measured detector information such as tracks and/or calorimeter cells into jet objects and by the parameters of the algorithm. 
The desired properties of such algorithms in $\pp(\bar{\mathrm{p}})$ collisions and in the corresponding theoretical framework have been discussed e.g. in \cite{Huth:1990mi}. 
In general, jet algorithms aim to reconstruct the kinematic properties of the initial parton with as little dependence on the details of its fragmentation process as possible, i.e. the algorithms should yield consistent results when applied in a theoretical calculation at any stage of a parton shower and at final state particle level.
A particularly well suited class of algorithms in this context are those using sequential recombination schemes, which are infrared and collinear safe, in contrast to many conceptually simpler cone algorithms.
The computationally optimized implementation of
sequential recombination algorithms in the FastJet package
\cite{Cacciari:2005hq} facilitates their applicability also in collision systems with high multiplicity and thereby the comparison of results obtained with the same jet
algorithms in $\pp$, $\pPb$, and $\PbPb$ collisions.
An additional complication in the context of jet reconstruction in high-multiplicity events arises from the large background particle density, i.e. particles in the same aperture as the jet that are not related to the initial hard scattering. 
This background can be subtracted on an event-by-event basis 
and the impact on the reconstructed jet observable needs to be evaluated carefully \cite{Cacciari:2011tm,Abelev:2013kqa,Abelev:2012ej}.

In this paper, jets reconstructed from charged particles  (\emph{charged jets}) with the
anti-$\kT$ algorithm measured with the ALICE detector in $\pPb$
collisions at $\snn = 5.02$ TeV are reported for different resolution
parameters, $R$. 
Section 2 describes in detail the correction steps needed in the analysis,
including the effect of the event background and its fluctuations on
the jet observables and the unfolding procedure to account for
background as well as detector effects. 
The results are presented and discussed in Sec. 3.

\section{Data Analysis}
\label{sec:techniques}
\subsection{Event and Track Selection}

The data used for this analysis were taken with the ALICE detector \cite{Abelev:2014ffa} during the $\pPb$ run of the LHC at $\sqrt{s_\mathrm{NN}} = 5.02 \TeV$ at the beginning of 2013.
Minimum bias events have been selected requiring at least one hit in both of the scintillator trigger detectors (V0A and V0C) covering the pseudorapidity $2.8 < \eta_\mathrm{lab} < 5.1$ and  $-3.7 < \eta_\mathrm{lab} < -1.7$, respectively \cite{Aamodt:2010pb}.
Here and in the following, $\eta_\mathrm{lab}$ denotes the
pseudorapidity in the ALICE laboratory frame.
Compared to this frame (with positive $\eta$ in the
direction of the V0A), the nucleon-nucleon center-of-mass system 
moves in rapidity by $y_\mathrm{NN} = -
0.465$ in the direction of the proton beam \cite{Adam:2014qja}. 

The event sample used in the analyses presented in this manuscript was collected exclusively for the beam configuration where the proton travels from V0A to V0C (clockwise). 
A van-der-Meer scan was used to measure the visible cross section $\sigma_{V0} = 2.09 \pm
0.07$~b for this case \cite{Abelev:2014epa}. 
Monte Carlo studies show that the sample consists mainly of non-single diffractive (NSD) interactions and a negligible contribution from single diffractive and electromagnetic interactions (for more details see \cite{Adam:2014qja,ALICE:2012xs}). 
The trigger is not fully efficient for NSD events. 
This inefficiency affects only events without a reconstructed vertex,
i.e. with no particles reconstructed within the acceptance of the SPD.  
The loss of efficiency is estimated to be
2.3\% with a large systematic uncertainty of 3.1\%
\cite{Adam:2014qja}. 
In this paper, the normalization to NSD events is only used for the construction of the nuclear modification factor.   

In addition to the trigger selection,  timing and vertex-quality cuts are used to suppress pile-up and bad quality events. 
The analysis
requires a reconstructed vertex, which is the case for
98.2\% of the events selected by the trigger. In addition, events with
a reconstructed vertex $|z| > 10~\mathrm{cm}$ along the beam axis are
rejected. In total, about 96M events are used for the analysis.

Charged particles are reconstructed as tracks in the Inner Tracking System (ITS) \cite{Aamodt:2010aa} and the Time Projection Chamber (TPC) which cover the full azimuth and $|\eta_\mathrm{lab}| < 0.9$ \cite{Alme:2010ke}. 
For tracks with reconstructed track points close to the vertex (from the two inner Silicon Pixel Detector (SPD) layers of the ITS),
 a momentum resolution of  0.8\% (3.8\%) for $\pt = 1$~GeV/$c$ (50 GeV/$c$) is reached \cite{Abelev:2014ffa}. 
The azimuthal distribution of these high quality tracks is not completely uniform due to inefficient regions in the SPD.
This can be compensated by considering in addition tracks \textit{without} reconstructed track points in the SPD. 
For those tracks, the primary vertex is used as an additional constraint in the track fitting to improve the momentum resolution. 
This approach yields a very uniform tracking efficiency within the acceptance, which is needed to avoid geometrical biases of the jet reconstruction algorithm 
caused by a non-uniform density of reconstructed  tracks.  % due to reconstruction inefficiencies.
The procedure is described in detail in the context of jet reconstruction with ALICE in $\PbPb$ events \cite{Abelev:2013kqa}. 
For the analyzed data, the additional tracks (without SPD track points) constitute approximately
4.3\% of the used track sample. 
Tracks with $\pt > 0.15 \mathrm{~GeV}/c$ and within a pseudorapidity interval $|\eta_\mathrm{lab}|<0.9$ are used as input to the jet reconstruction.
The overall efficiency for charged particle detection, including the effect of tracking efficiency as well as the geometrical acceptance, is 70\% at $\pt = 0.15 \mbox{~GeV}/c$ and
increases to 85\% at $\pt = 1 \mbox{~GeV}/c$ and above.

\subsection{Jet reconstruction and background corrections}

For the present analysis, the anti-$k_\mathrm{T}$ algorithm from the FastJet package \cite{Cacciari:2008gp} has been used to reconstruct jets from measured tracks with resolution parameters of $R=0.2$ and $R=0.4$.
In general, jets are only considered for further analysis if the jet-axis is separated from the edge of the track acceptance in $\eta_\mathrm{lab}$ by at least the resolution parameter $R$ used in the jet finding, e.g. jets reconstructed with $R=0.4$ are accepted within $|\eta_\mathrm{jet,\,lab}| < 0.9 - 0.4 = 0.5$. 
The jet transverse momentum is calculated by FastJet using the $\pt$ recombination scheme. 
To enable background corrections, the area $A$ for each jet is determined internally by distributing \emph{ghost particles} into the area that is clustered  \cite{Cacciari:2008gn}. 
Ghost particles have vanishing momentum and therefore do not influence the jet finding procedure. 
By construction, the number of ghost particles in a jet is a direct measure for the jet area. 
%
% The area resolution is driven by the size that is assigned to the ghost particles and therefore their density. 
%
A ghost particle density of 200 per unit area (0.005 area per ghost particle) was used to obtain a good area resolution with a reasonable computing time.

\begin{figure}[ht]
  \subfigure{\includegraphics[width=0.99\linewidth]{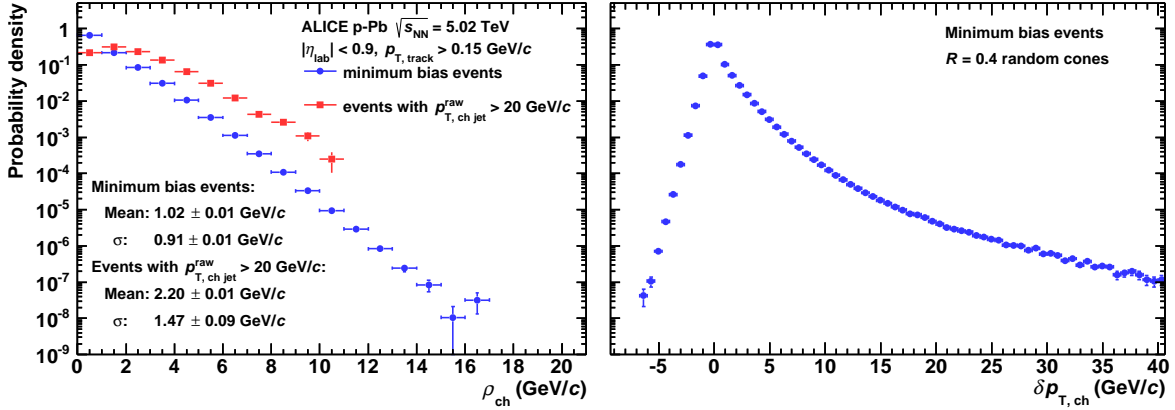} }\hfill
\caption{(Color online) Left: Probability distribution of the event-by-event transverse momentum background density (see Eq.~\ref{def:CMSBackground}).  The mean and variance for two event classes are indicated in the figure. $p_\mathrm{T,\,ch\;jet}^\mathrm{raw}$ represents uncorrected jet $\pT$. Right: Probability distribution of background fluctuations calculated with the random cone approach and defined via Eq.~\ref{def:DeltaPt} (resolution parameter $R=0.4$).}
\label{BackgroundPlots}
\end{figure}

In $\PbPb$ collisions, the background from particles not from the same hard scattering as the jet has a significant impact on the reconstructed jet momentum \cite{Abelev:2013kqa,Abelev:2012ej}.
The transverse momentum density of this background is estimated with a statistically robust method by using the median of all jet $\ptch$ per area  within one event for jets reconstructed with the $\kt$  algorithm.
In $\pPb$ collisions, the multiplicity density is two orders of magnitude smaller
than in central $\PbPb$ collisions \cite{ALICE:2012xs}, so a corresponding
reduction of the jet background is expected. To obtain a reliable
estimate for the more sparse environment of $\pPb$ events a modified
version of the approach described in \cite{Chatrchyan:2012tt} for
$\pp$ collisions is employed.
It uses the same method as in $\PbPb$, but contains an additional
correction factor, $C$, to account for regions without particles, which otherwise would not contribute to the overall area estimate.
The background density for each event is then given by
\eq{def:CMSBackground}{\rho_\mathrm{ch} = \mathrm{median} \left\{ \frac{p_{\mathrm{T},\,i}}{A_i} \right\}  \cdot C,}
where $i$ runs over all reconstructed $\kt$ jets in the event with
momentum $p_{\mathrm{T},\,i}$ and area $A_i$. $C$ is defined by
%\eq{def:BackgroundOccupancy}{C = \frac{\tn{Area of all } k_\mathrm{T} \tn{ jets containing tracks}}{\tn{Acceptance}}.}
\eq{def:BackgroundOccupancy}{C = \frac{\sum_j {A_{j,\,{k_\mathrm{T}}} }}{A_\tn{acc}}.}
Here, the numerator is the area of all $k_\mathrm{T}$ jets containing tracks and the denominator, $A_\tn{acc}$, is the acceptance in which charged particles are considered as input to the jet finding ($2 \times 0.9 \times 2\pi$).
The probability distribution for the background density in this
method, with the same track selection
criteria as the signal jet reconstruction and a radius of 0.4,  is shown in
Fig.~\ref{BackgroundPlots} (left).
The background density obtained with $R = 0.4$ is used both for
the correction of signal jets  with $R = 0.4$ and $R = 0.2$ to avoid
event-by-event fluctuations in the difference of the momenta for the two radii.
The probability distribution of  $\rho_\mathrm{ch}$ decreases approximately exponentially. It is smaller than $4$ GeV/$c$ for $98.6$\% of all events.
The mean background density and its variance for all events is $\langle\rho_\mathrm{ch}\rangle = 1.02$ GeV/$c$ (with negligible statistical uncertainty) and  $\sigma(\rho_\mathrm{ch}) = 0.91 \pm 0.01$ GeV/$c$. 
For events containing a jet with uncorrected transverse momentum $\ptch > 20$ GeV/$c$, it is $\langle\rho_\mathrm{ch}\rangle = 2.2 \pm 0.01$ GeV/$c$ and  $\sigma(\rho_\mathrm{ch}) = 1.47 \pm 0.09$ GeV/$c$, respectively.
The observed increase of the underlying event activity for events that contain a high-$\pt$ jet is expected. 
This increase is already present in $\pp$ collisions and has been quantified in detail  and with more differential  observables than the background density, e.g. in \cite{ALICE:2011ac}.  

The background density estimate provides an event-by-event correction for each jet with reconstructed transverse momentum $p_\mathrm{T,\,ch\;jet}$ and jet area $A_\mathrm{ch\;jet}$:
\begin{equation}
p_\mathrm{T,\,ch\;jet} = p_\mathrm{T,\,ch\;jet}^\mathrm{raw} -
A_\mathrm{ch\;jet} \cdot \rho_\mathrm{ch}.
\end{equation}
However, this approach neglects that the background for a given event is not uniformly distributed in the ($\eta_\mathrm{lab},\varphi$)-plane but fluctuates from region to region. 
These fluctuations are mainly Poissonian, but also encode correlated region-to-region variations of the particle multiplicity and the mean $\pT$ \cite{Abelev:2012ej}.
The effect of these fluctuations can be accounted for on a statistical basis in the unfolding of the measured jet $\ptch$-distributions.
The distribution of region-to-region density fluctuations around the event-wise background density estimate can be evaluated for the full event sample by a \textit{Random Cone} (RC) approach as described in \cite{Abelev:2012ej}. 
Cones with a radius $R$ corresponding to the resolution parameter of the jet finding algorithm are placed randomly in the ($\eta_\mathrm{lab},\varphi$) jet-acceptance and the transverse momenta for all tracks (charged particles) falling into this cone are summed and compared to the background estimate:
\eq{def:DeltaPt}{\deltaptch =
  \sum_\mathrm{i}{p_\mathrm{T,\,i}-\rho_\mathrm{ch} A}, ~~~ A = \pi R^2.}
The distribution of the residuals, $\deltaptch$, as shown in
Fig.~\ref{BackgroundPlots} (right) for $R = 0.4$, is a direct measure for
all intra-event fluctuations of the background and can be used
directly in the unfolding procedure.
In Fig.~\ref{BackgroundPlots} (right), a clear asymmetry of the distribution is visible.
It is caused by the fact that the $\deltaptch$ distribution of single particles sampled in the cone is asymmetric. 
Since the number of particles within a cone increases with its size, statistical fluctuations of the background estimate also increase (see also \cite{Abelev:2012ej}). 
Furthermore, the randomly placed cones can also overlap with jets.
In $\pPb$ collisions, there is the possibility for multiple hard collisions within
one $\pPb$ event, so a jet can also be the background to a jet from
another hard collision and contribute as an upward fluctuation. 
%!TEX encoding = UTF-8 Unicode
Therefore, an overlap of random cones with possible signal jets should not be \textit{a-priori} excluded in the fluctuation estimate, but is part of its systematic uncertainty. 
\subsection{Detector effects and unfolding}
\label{sec:DetectorEffects}

\begin{figure}[!htp]
\centering
\subfigure{\includegraphics[width=0.99\textwidth]{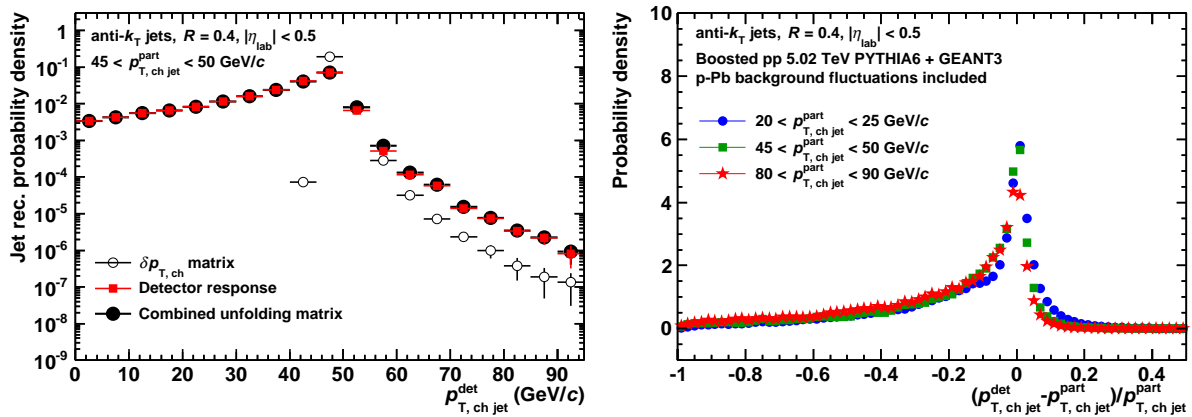}}\hfill
\caption{(Color online) Left: Projection of the combined unfolding matrix for jets with particle-level momentum $45 < p_\mathrm{T,\,ch\;jet}^\mathrm{part} < 50$ GeV/$c$. The matrix is obtained from the combination of  the detector response and background fluctuation matrices, which are also shown as projections (see text for details).
Right: Probability distribution of the relative difference between
particle-level (generated true) and detector-level charged jet transverse momentum for jets with different momenta. The effect of  background fluctuations is included for the jets reconstructed at detector level. Characteristic values of the distributions are summarized in Table~\ref{tab:residuals}. }
% used for the detector response matrix.
\label{DetectorResponsePlot}
\end{figure}

The main detector-related effects on the reconstructed jet are the reconstruction efficiency and the momentum resolution for single charged particles.
To determine the correction for these, a full detector simulation of $\pp$ jet events generated with PYTHIA6 (Perugia 2011, version 6.425) \cite{Sjostrand:2006za}  and GEANT3 particle transport \cite{Bru93a} is performed. 
In the simulation, two jet collections are matched geometrically (closeness in ($\eta_\mathrm{lab},\varphi$)-plane) with a one-to-one correspondence \cite{Abelev:2013kqa}: 
jets reconstructed at the charged particle level (\emph{part}) without detector effects and jets reconstructed from tracks after particle transport through the ALICE detector (\emph{det}).
In the simulation, the particle level reconstruction includes charged primary particles produced in the collision with $\pt > 0.15$\,GeV/$c$. 
Charged decay products from primary particle decays, excluding those from weak decays of strange particles, are included with the same $\pt$ threshold.
The response matrix is populated with matched particle- and detector-level 
jets. 
It relates the particle-level to the detector-level charged jet momentum and encodes the effects of single-particle momentum resolution and reconstruction efficiency on the reconstructed jet momentum.
A correction for the missing energy of neutral jet-constituents is not applied. 
The response is shown on a logarithmic scale in
Fig.~\ref{DetectorResponsePlot} (left) for charged jets with $R = 0.4$ and particle-level momentum between $45 < p_\mathrm{T,\,ch\;jet}^\mathrm{part} < 50$ GeV/$c$.
It can be seen that the most probable value for the reconstructed momentum is the particle-level momentum, but the distribution has large tails to the left and right. 
It is more probable that jets are reconstructed with a lower momentum than the truth, which is due to the dominating effect of the single-particle reconstruction efficiency that reduces the number of reconstructed particles in a jet. 
The tail to the right-hand side is mainly due to the single-particle momentum resolution, where a fraction of tracks is reconstructed with higher momentum than the truth, causing an upward shift of the jet momentum.

In addition, Fig.~\ref{DetectorResponsePlot} (left) shows the effect of the background fluctuations on the reconstructed jet momentum and the combination of detector effects and background fluctuations.
Even though the background fluctuations show a strong tail to the right-hand side, it is seen that in the combined unfolding matrix the effects of single-particle momentum resolution play the dominant role in reconstructing a jet with momentum higher than the truth. 
The default algorithm for the unfolding of the measured jet spectrum is based on the \textit{Singular Value Decomposition} (SVD) approach \cite{Hocker:1995kb} as implemented in the RooUnfold package \cite{Adye:2011gm}.
The default prior in the unfolding procedure is a smoothed version of the uncorrected jet spectrum itself.
In addition to the SVD unfolding approach, Bayesian
\cite{Blobel:2002pu,Dagostini:2003} and $\chi^2$ \cite{Blobel1984}
unfolding have been used for systematic comparisons and  validity
checks. 
The unfolded spectrum is also corrected for
unmatched jets using a jet reconstruction efficiency obtained from
generated--reconstructed comparison. 
This jet reconstruction efficiency is larger than 96\% in the considered momentum range.\\

\begin{table}[!htp]
\centering
\begin{tabular}{rr@{$\,\pm$\,}lr@{$\,\pm$\,}lr@{$\,\pm$\,}l}

 $p_\mathrm{T,\,ch\;jet}^\mathrm{part}$  & \multicolumn{2}{c}{$20-25 \mbox{~GeV/}c$} & \multicolumn{2}{c}{$45-50 \mbox{~GeV/}c$} & \multicolumn{2}{c}{$80-90 \mbox{~GeV/}c$} \\
 \hline
 MPV (Gaussian Fit) & $0.006$ & $0.002$ &  $-0.001$ & $0.002$ & $-0.010$ & $0.004$ \\
                  id. w/o bkg. fluct.             &  $0.007$ &$0.001$ & $-0.003$ &
                               $0.002$ & $-0.013$ & $0.004$ \\
                               Mean 
                               &  $0.149$ &$0.030$ &  $-0.181$ &
                               $0.030$ & $-0.222$ & $0.030$ \\
id. w/o bkg. fluct.                               & $-0.163$ & $0.030$ & $-0.188$ & $0.030$
                               & $-0.226$ & $0.030$ \\
                               Width $\sigma$ 
                               &  $0.238$ &$0.030$ &  $0.246$ &
                               $0.030$ & $0.259$ & $0.030$ \\
           id. w/o bkg. fluct.                    & $0.233$ & $0.009$ & $0.245$ & $0.005$  &
                               $0.258$ & $0.003$ \\

Quartile, 25\% above 
& $0.01$ & $0.01$ & $-0.01$ & $0.02$ & $-0.01$ & $0.02$ \\
id. w/o bkg. fluct.& $0.01$ & $0.01$ & $-0.01$ & $0.02$ & $-0.03$ & $0.02$ \\
Quartile, 50\% above 
& $-0.05$ & $0.04$ & $-0.09$ & $0.01$  & $-0.13$ & $0.04$ \\
id. w/o bkg. fluct. & $-0.07$ & $0.04$ & $-0.09$ & $0.04$  & $-0.13$ & $0.04$ \\
Quartile, 75\% above 
& $-0.25$ &  $0.06$ & $-0.29$ & $0.05$ & $-0.37$ & $0.04$ \\
id. w/o bkg. fluct.& $-0.27$ &  $0.04$ & $-0.29$ & $0.06$ & $-0.37$ & $0.04$ \\
\hline
\end{tabular} 
\caption{
\label{tab:residuals}
Characteristic values for the distribution of residuals of the total
charged jet response shown in Fig.~\ref{DetectorResponsePlot} (right), including the effect of
background fluctuations and without:
most probable value (MPV) determined via a Gaussian fit to the central
peak region, first and second moment (mean and width $\sigma$), and
quartiles. The precision of the quartiles is limited by the finite bin
width of 0.01.}
\end{table}

The influence of these detector effects and background
fluctuations on the jet momentum is shown for three transverse
momentum intervals in  Fig.~\ref{DetectorResponsePlot} (right) via the probability distribution of the relative difference of the detector-level and particle-level charged jet transverse momentum. 
For all momentum bins the distribution is asymmetric.  
The most probable response was determined using Gaussian fits to the peak region.
It can be seen in Tab.~\ref{tab:residuals} that it is close to zero ($\le 1\%$) with a mild $\pt$ dependence. 
%$p_\mathrm{T,\,ch\;jet}^\mathrm{det}$ is close to $p_\mathrm{T,\,ch\;jet}^\mathrm{part}$  
%
To further quantify the distributions, numerical values for their mean
and width are also given in Tab.~\ref{tab:residuals}. 
Since the width is not a well-defined measure of the jet momentum
resolution for these asymmetric distributions,  the quartiles of the
distribution are provided in addition. 
Approximately, 25\% of the jets  have a larger momentum than the generated. 
The 50\% (median) correction is only 5\% for  $p_\mathrm{T,\,ch\;jet}^\mathrm{part} = 20-25$\,GeV/$c$ and increases towards larger jet momenta. 
In Tab.~\ref{tab:residuals} the values for the respective
distributions without background fluctuations are also given (not shown in
Fig.~\ref{DetectorResponsePlot}).
Clearly, the instrumental response dominates the jet response as
already seen in Fig.~\ref{DetectorResponsePlot} (left).
The main effect of the background fluctuations is a broadening of the jet
response and an upward shift of the average reconstructed energy due
to the asymmetric shape of the fluctuations as seen in
Fig.~\ref{BackgroundPlots} (right). The most probable value remains
unaffected within the uncertainties when background fluctuations are included. 

\subsection{Nuclear modification factor}
The nuclear modification factor compares a $\pt$-differential yield
in $\pPb$ collisions to the differential production cross section in $\pp$ collisions at the 
same $\snn$ to quantify nuclear effects:
\begin{equation}
\label{eq:r_ppb}
\RpPb =
\frac{
\mathrm{d^2}N_\mathrm{pPb}/\mathrm{d}\eta\mathrm{d}p_\mathrm{T}}
{\left<T_\mathrm{pPb}\right>   \cdot \mathrm{d^2}\sigma_\mathrm{pp}/\mathrm{d}\eta\mathrm{d}p_\mathrm{T}}.
\end{equation}
Here, $\left<T_\mathrm{pPb}\right>$ is the nuclear overlap 
function which accounts for the increased parton flux in $\pPb$ compared to $\pp$ collisions.  
It is related to the number of binary nucleon-nucleon collisions via $\left<T_\mathrm{pPb}\right> =
\left<N_\mathrm{coll}\right> / \sigma_\mathrm{INEL}^\mathrm{pp}$ and has been calculated
in a Glauber Monte Carlo, as described in \cite{Adam:2014qja}.
Here, $\sigma_\mathrm{INEL}^\mathrm{pp}$ represents the total inelastic cross section in $\pp$ collisions. 
For minimum bias $\pPb$ collisions, the nuclear overlap function is $\left<T_\mathrm{pPb}\right> = (0.0983 \pm
0.0034)\,\mathrm{mb}^{-1}$ and $\left<N_\mathrm{coll}\right> = 6.87 \pm 0.56$. 
In this paper, the reference differential production cross section in $\pp$  is constructed from
the ALICE  charged jet measurement at 7~TeV \cite{ALICE:2014dla} by a pQCD based scaling.
In the nuclear modification factor, the invariant yield for NSD events  in $\pPb$ is compared to inelastic $\pp$ collisions. 
Hence, the additional correction of $(2.3 \pm 3.1)\%$ is applied as discussed above.

\subsection{NLO calculations and $\pp$ reference}
\label{sec:ppreference}

Perturbative QCD calculations are used for two purposes in this paper: for comparison to the measurement of jet production in $\pPb$, and  as additional input to the construction of the $\pp$ reference.
The calculations have been performed within the POWHEG box framework \cite{Nason:2004rx,Frixione:2007nw}, which facilitates next-to-leading order (NLO) precision in calculating parton scattering cross sections in an event-by-event Monte Carlo. 
Event-by-event the outgoing partons from POWHEG are passed to PYTHIA8 \cite{Sjostrand:2007gs} where the subsequent parton shower is handled. 
For this, a POWHEG version matched to the PYTHIA8 fragmentation is used to avoid double counting of NLO effects already considered in the PYTHIA8 code.
The Monte Carlo approach has the advantage that the same selection criteria and jet finding algorithm can be used on final state particle level, as in the analysis of the real data, in particular, the limitation to charged constituents of a jet.
The dominant uncertainty in the parton level calculation is given by the choice of renormalization scale, $\mu_\mathrm{R}$, and factorization scale, $\mu_\mathrm{F}$. 
The default value has been chosen to be $\mu_\mathrm{R} = \mu_\mathrm{F} = \pt$ and independent variations by a factor of two around the central value are considered as the systematic uncertainty. 
% scale 13%, CTEQ6% EPS9%
In addition, the uncertainty on the parton distribution functions has been taken into account by the variation of the final results for the respective error sets of the parton density functions (PDFs).

For the comparison with the measured $\pPb$ data, proton PDFs corrected for nuclear effects (CTEQ6.6 \cite{Nadolsky:2008zw} with EPS09 \cite{Eskola:2009uj}) have been used. 
Prior to passing the scattered partons to PYTHIA8 for showering, they can be boosted into the same reference frame as the $\pPb$ reaction by $y_\mathrm{NN} = 0.465$.

The construction of the  $\pp$ reference at $\snn =
5.02$~TeV is based on the ALICE measurement of charged jets in $\pp$ collisions at 7~TeV, 
described in detail in \cite{ALICE:2014dla}.
For the purpose of the reference scaling, the same analysis chain has been
used as for $\pPb$. The same binning in pseudorapidity and transverse
momentum allows for a partial cancellation of common systematic uncertainties in the $\pp$ and $\pPb$ data sets.
In addition, the same background subtraction approach as in the $\pPb$ analysis is used for the $\pp$ data.
%
% The resulting spectra are fully compatible with the published data.
%
In the present analysis, the scaling is done with a factor which is 
determined for
each $\ptch$ bin by the NLO pQCD calculations  (POWHEG+PYTHIA8) at the two energies.
For the $\pp$ reference scaling the parton distribution functions in the POWHEG calculation have been replaced by the free proton PDF from CTEQ6.6.
The scaling factor is given by
 \begin{equation}
\label{eq:ppRefScaling}
F (\pt) = \frac{\left.\mathrm{yield}(\ptch)\right|^{5.02
    \TeV,\, \mathrm{boosted}}_\mathrm{pp,\, NLO}}{\left.\mathrm{yield}(\ptch)\right|^{7
    \TeV}_\mathrm{pp,\, NLO}}.
 \end{equation}
The factor decreases monotonically from $F \approx 0.65$ to $0.45$ in the reported $\pT$ range.
As already described above, the laboratory frame is not the center-of-mass frame of the
collision as is the case for $\pp$ collisions. 
Therefore, the numerator of the
scaling factor $F$ in Eq.~\ref{eq:ppRefScaling} is determined in the NLO
calculation where the additional Lorentz-boost is applied to the hard
scattered partons prior to fragmentation. 
The resulting reduction of the observed jets for $|\eta_\mathrm{lab}|
<0.5$  is smaller than 5\% in the relevant momentum range. 

\subsection{Jet production cross section ratio}

The broadening or narrowing of the parton shower with respect to the original parton direction can have a direct impact on the jet production cross section reconstructed with different resolution parameters. This can be tested via the ratio of yields or cross sections in common rapidity interval, here $|\eta_\mathrm{lab}| < 0.5$ for $ R = 0.2$ and $0.4$:
\begin{equation}
\label{eq:xsec_ratio}
\mathscr{R}(0.2,\,0.4) = \frac{\mathrm{d}\sigma_\mathrm{pPb,\, R=0.2} / \mathrm{d}p_\mathrm{T}}{\mathrm{d}\sigma_\mathrm{pPb,\, R=0.4} / \mathrm{d}p_\mathrm{T}}.
\end{equation}
Considering the extreme scenario that all fragments are already contained within $R = 0.2$ this ratio is unity. 
In this case, also the statistical uncertainties between $R = 0.2$ and $R  = 0.4$ are fully correlated and cancel completely in the ratio, when the jets are reconstructed from the same data set. 
In the case the jets are less collimated, the ratio decreases and the statistical uncertainties only cancel partially. 
For the analysis presented in this paper, the conditional probability for reconstructing a $R =0.2$ jet in the same $\pt$-bin as a geometrically close $R = 0.4$ jet is $25 - 50$\%, which leads to a reduction of the statistical uncertainty of the ratio of $5-10\%$ compared to the case of no correlation.

\subsection{Systematic uncertainties}
\label{sec:uncertainties}
The various sources of systematic uncertainties are listed in Tab.~\ref{tab:sys_err} for the full $\pt$-range of the three observables presented in this paper: jet production cross section, nuclear modification factor, and cross section ratio.
The most important sources will be discussed in the following.

The dominant source of uncertainty for the $\pt$-differential jet production
cross section is the imperfect knowledge of the single-particle tracking
efficiency that has a direct impact on the correction of the jet
momentum in the unfolding, as discussed above.
In $\pPb$ collisions, the single-particle efficiency is known with a relative accuracy of 4\%, which is equivalent to a 4\% uncertainty on the jet momentum scale.
To estimate the effect of the tracking efficiency uncertainty on the jet yield, the tracking efficiency is artificially lowered
by randomly discarding a certain fraction (4\% in $\pPb$) of tracks used as input for the jet finding. 
Depending on the shape of the spectrum, the uncertainty on the single particle efficiency (jet momentum scale) translates into an uncertainty of  8 to 15\% on the yield.

To estimate the uncertainty on the $\pPb$ nuclear modification factor,
the uncertainty on the single-particle tracking efficiency in the two collision systems ($\pp$ and $\pPb$) has to be evaluated.
This uncertainty on the efficiency is correlated between the data sets,
since the correction is determined with the same underlying Monte
Carlo description of the ALICE detector and for similar track
quality cuts.
Only variations of  detector conditions  between run
periods may  reduce the degree of correlation.
The uncorrelated uncertainty has been estimated to be 2\%, and the uncertainty for the nuclear modification factor has been determined by artificially introducing such a difference in the tracking efficiency between the two collision systems.  

The uncertainty on the spectra induced by the underlying event
subtraction has been estimated by comparing the results with various methods for background subtraction; ranging from
purely track-based to jet-based density estimates, including an $\eta_\mathrm{lab}$-dependent correction. 
As seen in Fig.~\ref{BackgroundPlots}, a typical correction $\pi R^2\rho_\mathrm{ch}$ for a jet with $R = 0.4$ is about 1 GeV/$c$. 
The uncertainty on this correction can be treated similar to an
uncertainty on the jet momentum scale. For the final spectrum, the uncertainty on the yields from the background correction method is approximately $2$\%.

In the determination of the fluctuations of the underlying event, the main uncertainty is given by the exclusion of reconstructed jets in the random cone sampling of the event. 
The probability for a random cone to overlap with reconstructed jets is higher than for the jets itself. On average, a jet can overlap with $N_\mathrm{coll}-1$ jets in one event. The random cone can overlap with $N_\mathrm{coll}$ jets. To account for this, the $\deltaptch$ calculation can be modified to discard on a statistical basis random cones that overlap with signal jets. This lowers the average overlap probability.
However, since this modified $\deltaptch$ calculation strongly depends on the signal jet definition and also on how an overlap is defined, it is not used by default but considered for systematic uncertainties.
The effect of this partial signal exclusion approach on the fully corrected jet yields is of the order of 0.1\%.

\begin{figure}[!htp]
\centering
\subfigure{\includegraphics[width=1.0\textwidth]{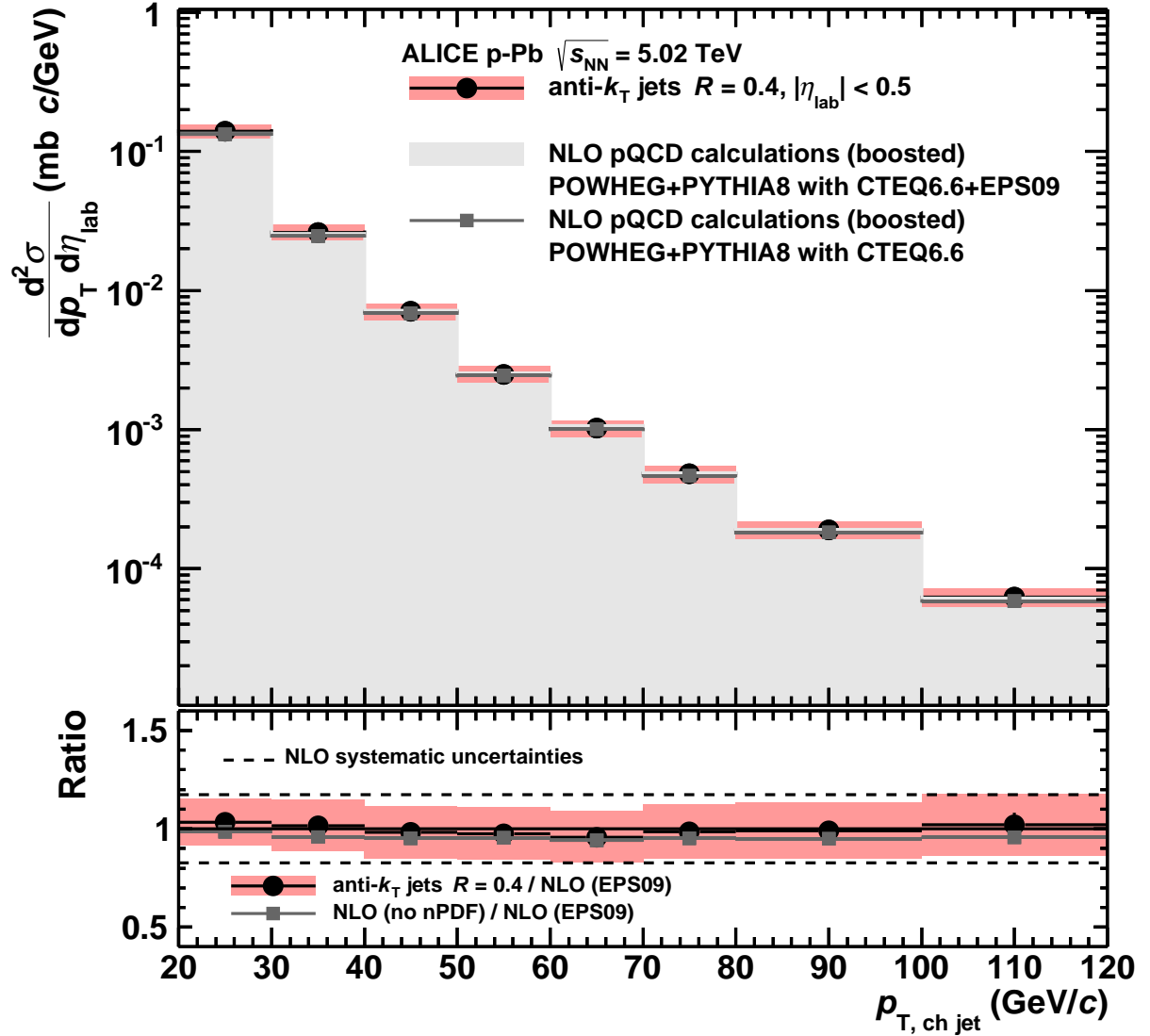}}\hfill
\caption{(Color online) Top panel: $\pt$-differential production cross section of charged jet production in $\pPb$ collisions at $5.02$ TeV for $R=0.4$. Bottom panel: Ratio of data and NLO pQCD calculations. The global uncertainty from the measurement of the visible cross section of $3.5\%$ is not shown. The uncertainties on the pQCD calculation are only shown in the ratio plot as dashed lines. The pQCD calculations take into account the rapidity shift of the nucleon-nucleon center-of-mass system in $\pPb$ with a boosted parton system.}
\label{FinalSpectrumR040}
\end{figure}

The uncertainty of the scaling procedure to obtain the reference
spectrum is estimated by determining the scaling factors F(\pt) after varying the scales $\mu_\mathrm{R}$ and $\mu_\mathrm{F}$ in the POWHEG
NLO generation, and by using different tunes in the outgoing
fragmentation handled by PYTHIA8. 
Furthermore, standalone calculations with PYTHIA6 and PYTHIA8 using
different generator tunes and with HERWIG at the two energies have
been performed to obtain scaling factors according to Eq.~\ref{eq:ppRefScaling}.
A general uncertainty for how well LO generators and NLO calculations can describe the $\sqrt{s}$-dependence of particle production is also considered: 
in ALICE measurements of the $\pi^0$ production in $\pp$ collisions, it has been observed that pQCD calculations predict  a stronger increase of the production cross section when going from $0.9$ to 7~TeV than supported by the data \cite{Abelev:2012cn}. 
A similar effect is also seen in unidentified charged hadrons measured with ALICE at 0.9, 2.76, and 7~TeV \cite{Abelev:2013ala}.
Furthermore, the $\sqrt{s}$-dependence of the jet production cross section has been cross checked internally with an interpolation between 7 and 2.76~TeV, using preliminary ALICE results on  charged jets at $\sqrt{s} = 2.76$~TeV.
In total, these studies yield an additional uncertainty on the $\pp$
reference of  $10\%$ for the extrapolation from 7 to 5.02 TeV. 
It is reported as an independent
normalization uncertainty, similar to the uncertainty on the nuclear
overlap function.

%%%%%%%%%%

\begin{table}
\begin{tabular}{lccccc}
Observable & \multicolumn{2}{c}{Jet cross section} & \multicolumn{2}{c}{$\RpPb$} & \multicolumn{1}{c}{$\mathscr{R}$}\\
Resolution parameter &$R=0.2$& $R=0.4$ & $R=0.2$ & $R=0.4$ & 0.2/0.4\\

%\multicolumn{1}{r}{$p_{\rm T}$-bin (GeV/$c$)}  &  $30-40$& $70-80$ &  $30-40$& $70-80$ &  $30-40$& $70-80$ &  $30-40$& $70-80$ &  $30-40$& $70-80$\\
Uncertainty source & \multicolumn{5}{c}{} \\ 

\hline
Single-particle efficiency (\%) & $7.9-12.8$ & $10.2-14.2$ & $4.1-5.9$ & $4.9-6.3$ & $2.1-2.1$ \\
Unfolding (\%) & 2.2 & 1.7 & 2.8 & 2.2 & 1.5 \\
Unfolding prior steepness (\%) & $1.4-4.8$ & $0.5-4.0$ & $2.9-8.0$ & $0.9-4.4$ & $1.1-1.5$ \\
Regularization strength (\%) & $3.1-3.9$ & $2.3-4.4$ & $3.6-5.8$ & $2.3-5.6$ & $1.1-4.7$ \\
Minimum $\pt$ cut-off (\%) & $1.1-0.3$ & $2.3-0.1$ & $1.3-1.4$ & $2.8-4.1$ & $1.2-0.4$ \\
Background estimate (\%) & $1.8-0.6$ & $3.7-1.5$ & $1.8-0.6$ & $3.7-1.5$ & $2.0-0.9$ \\
$\deltaptch$ estimate (\%) & $0.0-0.0$ & $0.1-0.0$ & $0.0-0.0$ & $0.1-0.0$ & $0.1-0.0$ \\
\hline
Combined uncertainty (\%) & $9.2-14.4$ & $11.5-15.5$ & $7.1-11.9$ & $7.5-10.7$ & $3.8-5.7$ \\
&&&&&\\
$\left<T_\mathrm{pPb}\right>$ (\%) & - & - & 3.4 & 3.4 & - \\
pp cross section (\%) & - & - & 3.5 & 3.5 & - \\
Reference scaling pp 7 TeV (\%) & - & - & 10.0 & 10.0 & - \\
NSD selection efficiency $\pPb$ (\%) & - & - & 3.1 & 3.1 & - \\
\hline
Combined scaling uncertainty (\%) & - & - & 11.6 & 11.6 & - \\

\end{tabular} 
\caption{
\label{tab:sys_err}
Summary of systematic uncertainties on the fully corrected jet spectrum, the corresponding nuclear modification factor, and the jet production cross section ratio for the resolution parameters $R=0.2$ and $R=0.4$. 
The percentages are given for the whole shown transverse momentum range $20-120$ GeV/$c$}
\end{table}

%%%%%%%%%%%%%%%%%%%%%%%%%%%
%%%%%%%%%%%%%%%%%%%%%%%%%%% RESULTS
%%%%%%%%%%%%%%%%%%%%%%%%%%%
 
\section{Results}
\label{sec:results}

\begin{figure}[!htp]
\centering
\subfigure{\includegraphics[width=1.0\textwidth]{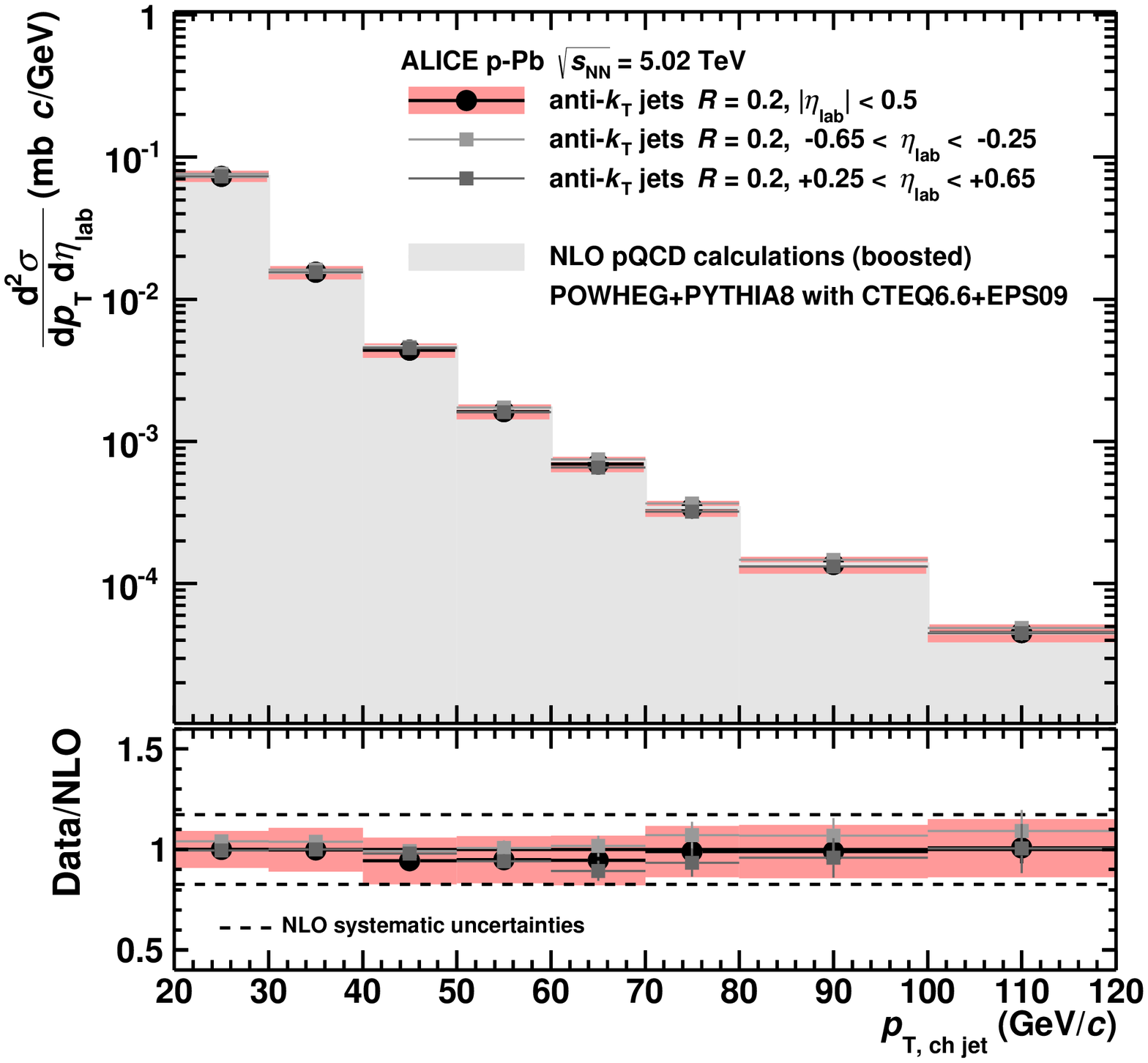}}\hfill
\caption{(Color online) Top panel: $\pt$-differential production cross section of charged jet production  in $\pPb$ collisions at $5.02$ TeV for $R=0.2$. Bottom panel: Ratio of data and NLO pQCD calculations. The global uncertainty from the measurement of the visible cross section of $3.5\%$ is not shown. The uncertainties on the pQCD calculation are only shown in the ratio plot as dashed lines. The pQCD calculations take into account the rapidity shift of the nucleon-nucleon center-of-mass system in $\pPb$ with a boosted parton system. }
\label{FinalSpectrumR020}
\end{figure}

The $\pt$-differential production cross sections for jets reconstructed from charged particles in minimum bias $\pPb$ collisions at $\snn = 5.02$~TeV 
are shown in Figs.~\ref{FinalSpectrumR040} and~\ref{FinalSpectrumR020} for the resolution parameters $R=0.4$ and $R=0.2$.
The spectra are found to agree well with scaled NLO pQCD calculations
(POWHEG+PYTHIA8) using nuclear PDFs (CTEQ6.6+EPS09) as seen best in the ratio data over calculation in the lower panels.  
However, the effect of the nuclear PDFs on the jet production in the reported kinematic regime is almost negligible, as seen in the comparison to calculations with only proton PDFs (CTEQ6.6).
Figure~\ref{FinalSpectrumR020} also shows the jet spectra for $-0.65 < \eta_\mathrm{lab} < -0.25$ and  $0.25 < \eta_\mathrm{lab} < 0.65$ compared to the results from the symmetric selection $|\eta_\mathrm{lab}| < 0.5$. 
Here, $\eta_\mathrm{lab}$ denotes the pseudorapidity of the jet axis.
The first selection roughly corresponds to a small window around mid-rapidity for the
nucleon-nucleon center-of-mass system, while the second is separated
from it by about one unit in rapidity.  
No significant change of the jet spectra is observed for these two
$\eta_\mathrm{lab}$ regions centered at $-0.45$ and $0.45$.
Thus, the jet measurement has no strong sensitivity to the
rapidity shift and the pseudorapidity dependent variation of the
multiplicity (underlying event)  within the statistical and systematic uncertainties of the measurement.

\begin{figure}[t!htp]
\centering
\subfigure{\includegraphics[width=1.0\textwidth]{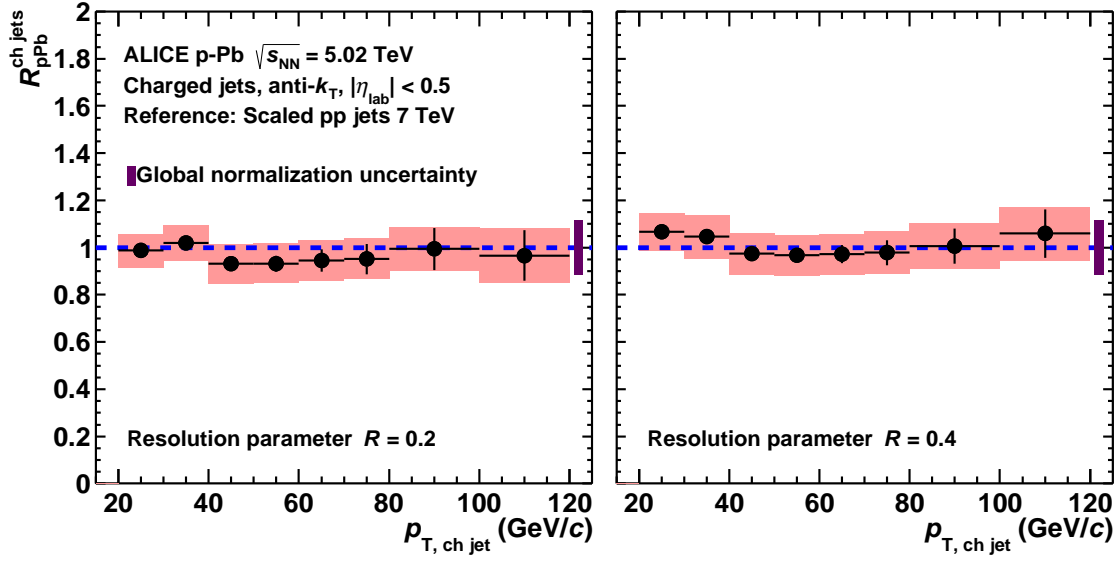}}\hfill
\caption{(Color online) Nuclear modification factors  $R_\mathrm{pPb}$ of charged
  jets for $R=0.2$ (left) and $R=0.4$ (right). The combined global normalization  uncertainty from $\left<T_\mathrm{pPb}\right>$, the correction to NSD events, the measured pp cross section, and the reference scaling is depicted by the box around unity.}
\label{FinalRpPb}
\end{figure}

The  nuclear modification factor $\RpPb$ is constructed based on the $\pt$-differential yields and the extrapolated $\pp$ production cross section at $5.02$ TeV for $R = 0.2$ and $0.4$. 
It is shown in the left and right panel of Fig.~\ref{FinalRpPb}, respectively. 
In the reported $\pt$-range, it is consistent with unity, indicating the absence of a large modification of the initial parton distributions or a strong final state effect on jet production.
Before comparing these results to the measured single-particle results for $\RpPb$, one has to consider that the same reconstructed $\pt$ corresponds to a different underlying parton transverse momentum.
Assuming that all spectra should obey the same power law behavior at high $\pt$, an effective conversion between the spectra can be derived at a given energy via the POWHEG+PYTHIA8 simulations described above. 
To match the single charged particle spectra in the simulation to charged jets with $R = 0.4$, a transformation $p_\mathrm{T}^{h^\pm} \rightarrow 2.28  p_\mathrm{T}^{h^\pm}$ is needed.
Thus, the reported nuclear modification factor for charged jets probes roughly the same parton $\pt$-region as the ALICE measurement of single charged
particles that shows a nuclear modification factor in agreement with unity
in the measured high-$\pt$ range up to 50 GeV/$c$ \cite{Abelev:2014dsa}. 

Since the jet measurements integrate the final state particles, they have a smaller sensitivity to the fragmentation pattern of partons than single particles.
Differences between the nuclear modification factor for jets and single high-$\pt$ particles, as suggested by measurements in \cite{Khachatryan:2015xaa,ATLAS-CONF-2014-029}, could point to a modified fragmentation pattern or differently biased jet selection in $\pPb$ collisions. 

\begin{figure}[thp]
\subfigure{\includegraphics[width=0.85\textwidth]{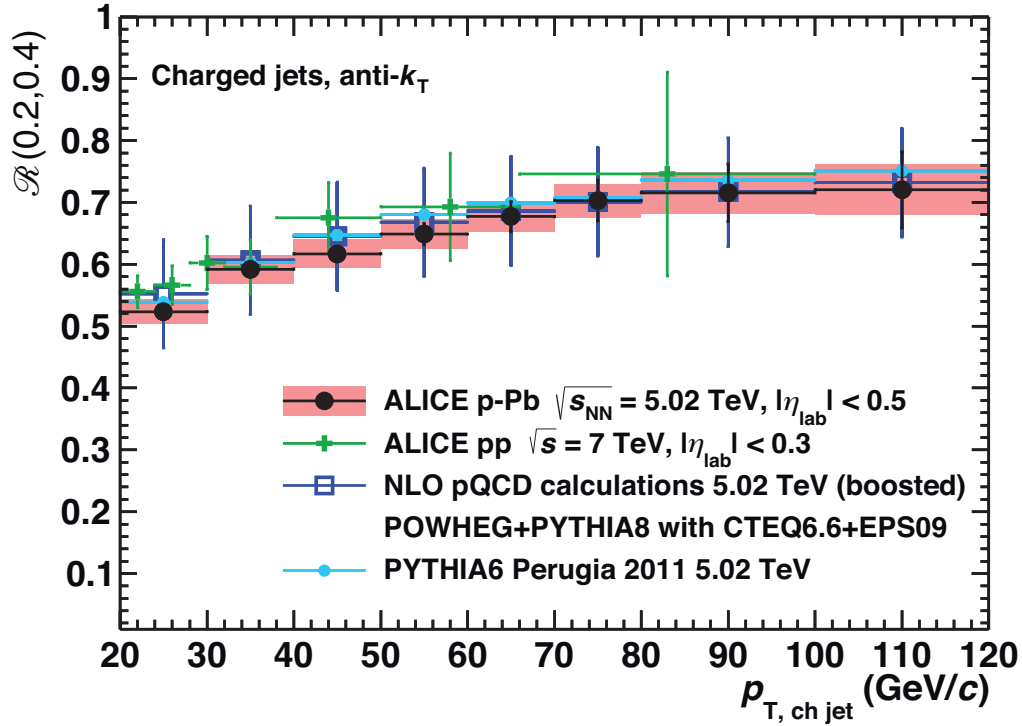}}\hfill
\caption{(Color online) Charged jet production cross section ratio for different resolution parameters as defined in Eq.~\ref{eq:xsec_ratio}. The data in $\pPb$ collisions at $\snn = 5.02 \TeV$ are compared to PYTHIA6 (tune: Perugia 2011, no uncertainties shown) and POWHEG+PYTHIA8 (combined stat. and syst. uncertainties shown) at the same energy, and to pp collisions at 7 TeV (only stat. uncertainties shown).}
\label{FinalShape}
\end{figure}

A modified fragmentation pattern may be also reflected in the collimation or transverse structure of jets. A first step in testing possible cold nuclear matter effects on the
jet structure is the ratio of jet production cross sections for two different resolution
parameters. 
It is shown for $R=0.2$ and $R=0.4$ in $\pPb$ in
Fig.~\ref{FinalShape} and compared to PYTHIA6 (Tune Perugia 2011) and 
POWHEG+PYTHIA8 at  $\snn = 5.02$ TeV and to ALICE results in $\pp$ collisions at $\sqrt{s} = 7$ TeV
\cite{ALICE:2014dla}. 
All data show the expected increase of the ratio from the increasing collimation of jets for higher transverse momentum and agree well within the uncertainties. No significant energy dependence or change with collision species is observed.
The data for $\pPb$ collisions is well described by the NLO calculation as well as by the simulation of $\pp$ collisions with PYTHIA6 at the same energy.  
It should be noted that the ratio for charged jets is, in general, above the ratio obtained for fully reconstructed jets, containing charged and neutral constituents. 
This can be understood from the contribution from neutral pions that decay already at the collision vertex and lead to an effective broadening of the jet profile when including the neutral component in the jet reconstruction, mainly in the form of decay photons.
For the same reason, the inclusion of the hadronization in the NLO pQCD calculation is essential to describe the ratio of jet production cross section as also discussed in \cite{Abelev:2013fn}.
%

%%%%%%%%%%%%%%%%%%%%%%%%%%%
%%%%%%%%%%%%%%%%%%%%%%%%%%% Summary
%%%%%%%%%%%%%%%%%%%%%%%%%%%

\section{Summary}
\label{sec:summary}

In this paper, $\pt$-differential charged jet production cross sections in $\pPb$ collisions at $\sqrt{s_\mathrm{NN}} = 5.02$ TeV have been shown up to $p_{\mathrm{T,\,ch\;jet}}$ of 120 $\mbox{~GeV/}c$ for resolution parameters $R=0.2$ and $R = 0.4$. The charged jet production is found to be compatible with scaled pQCD calculations at the same energy using nuclear PDFs.
At the same time, the nuclear modification factor $R_\mathrm{pPb}$ (using a scaled measurement of jets in $\pp$ collisions at $\sqrt{s} = 7$ TeV as a reference) does not  show strong nuclear effects on jet production and is consistent with unity for $R = 0.4$ and $R=0.2$ in the measured $\pt$-range between 20 and 120 GeV/$c$.
The jet cross section ratio of $R=0.2/0.4$ is compatible with 7 TeV $\pp$ data and also with the predictions from PYTHIA6 Perugia 2011 and POWHEG+PYTHIA8 calculations at  5.02 TeV. 
No indication of a strong nuclear modification of the jet radial profile is observed, comparing jets with different resolution parameters $R=0.2$ and $R=0.4$.

%%%%%%%% acknowledgements
\newenvironment{acknowledgement}{\relax}{\relax}
\begin{acknowledgement}
\section*{Acknowledgements}
\input{acknowledgements_jan2015.tex}    %%%%%%% get the lates version before submitting
\end{acknowledgement}
%
%
%%%%%%%% Bibliography (In case of using bibtex generate the bbl requested by arXiv)
%\bibliographystyle{style}   % Put here the style file name for the paper, i.e.apsrev4-1
%\bibliography{biblio}
%\input {bibliography.tex}  
%

% \bibliographystyle{mybibstyle}
\bibliographystyle{utphys}
\bibliography{paper}

\newpage
%
%\input{}               %%%%%%%%%%% put your appendices here
%
%%%%%%%%% appendix with author list
\appendix
\section{The ALICE Collaboration}
\label{app:collab}
\input{Alice_Authorlist_2015-Feb-18.tex}  %%%%%%% done by webmaster team
\end{document}

%% file: helpers.tex
\newcommand{\tn}[1]{\textnormal{#1}}
\newcommand{\eq}[2]{\begin{equation}\label{#1} #2 \end{equation}}

\newcommand{\TeV}[0]{\tn{ TeV}}
\newcommand{\GeV}[0]{\tn{ GeV}}

\usepackage{float}

%% file: acknowledgements_jan2015.tex
The ALICE Collaboration would like to thank all its engineers and technicians for their invaluable contributions to the construction of the experiment and the CERN accelerator teams for the outstanding performance of the LHC complex.
The ALICE Collaboration gratefully acknowledges the resources and support provided by all Grid centres and the Worldwide LHC Computing Grid (WLCG) collaboration.
The ALICE Collaboration acknowledges the following funding agencies for their support in building and
running the ALICE detector:
State Committee of Science,  World Federation of Scientists (WFS)
and Swiss Fonds Kidagan, Armenia,
Conselho Nacional de Desenvolvimento Cient\'{\i}fico e Tecnol\'{o}gico (CNPq), Financiadora de Estudos e Projetos (FINEP),
Funda\c{c}\~{a}o de Amparo \`{a} Pesquisa do Estado de S\~{a}o Paulo (FAPESP);
National Natural Science Foundation of China (NSFC), the Chinese Ministry of Education (CMOE)
and the Ministry of Science and Technology of China (MSTC);
Ministry of Education and Youth of the Czech Republic;
Danish Natural Science Research Council, the Carlsberg Foundation and the Danish National Research Foundation;
The European Research Council under the European Community's Seventh Framework Programme;
Helsinki Institute of Physics and the Academy of Finland;
French CNRS-IN2P3, the `Region Pays de Loire', `Region Alsace', `Region Auvergne' and CEA, France;
German Bundesministerium fur Bildung, Wissenschaft, Forschung und Technologie (BMBF) and the Helmholtz Association;
General Secretariat for Research and Technology, Ministry of
Development, Greece;
Hungarian Orszagos Tudomanyos Kutatasi Alappgrammok (OTKA) and National Office for Research and Technology (NKTH);
Department of Atomic Energy and Department of Science and Technology of the Government of India;
Istituto Nazionale di Fisica Nucleare (INFN) and Centro Fermi -
Museo Storico della Fisica e Centro Studi e Ricerche "Enrico
Fermi", Italy;
MEXT Grant-in-Aid for Specially Promoted Research, Ja\-pan;
Joint Institute for Nuclear Research, Dubna;
National Research Foundation of Korea (NRF);
Consejo Nacional de Cienca y Tecnologia (CONACYT), Direccion General de Asuntos del Personal Academico(DGAPA), M\'{e}xico, :Amerique Latine Formation academique – European Commission(ALFA-EC) and the EPLANET Program
(European Particle Physics Latin American Network)
Stichting voor Fundamenteel Onderzoek der Materie (FOM) and the Nederlandse Organisatie voor Wetenschappelijk Onderzoek (NWO), Netherlands;
Research Council of Norway (NFR);
National Science Centre, Poland;
Ministry of National Education/Institute for Atomic Physics and Consiliul Naţional al Cercetării Ştiinţifice - Executive Agency for Higher Education Research Development and Innovation Funding (CNCS-UEFISCDI) - Romania;
Ministry of Education and Science of Russian Federation, Russian
Academy of Sciences, Russian Federal Agency of Atomic Energy,
Russian Federal Agency for Science and Innovations and The Russian
Foundation for Basic Research;
Ministry of Education of Slovakia;
Department of Science and Technology, South Africa;
Centro de Investigaciones Energeticas, Medioambientales y Tecnologicas (CIEMAT), E-Infrastructure shared between Europe and Latin America (EELA), Ministerio de Econom\'{i}a y Competitividad (MINECO) of Spain, Xunta de Galicia (Conseller\'{\i}a de Educaci\'{o}n),
Centro de Aplicaciones Tecnológicas y Desarrollo Nuclear (CEA\-DEN), Cubaenerg\'{\i}a, Cuba, and IAEA (International Atomic Energy Agency);
Swedish Research Council (VR) and Knut $\&$ Alice Wallenberg
Foundation (KAW);
Ukraine Ministry of Education and Science;
United Kingdom Science and Technology Facilities Council (STFC);
The United States Department of Energy, the United States National
Science Foundation, the State of Texas, and the State of Ohio;
Ministry of Science, Education and Sports of Croatia and  Unity through Knowledge Fund, Croatia.
Council of Scientific and Industrial Research (CSIR), New Delhi, India

%% file: Alice_Authorlist_2015-Feb-18.tex
% Collaboration: CERN-LHC-ALICE
% Generation Date is 2015/Feb/18

% How to use:
%%%%%%%%% appendix with author list
%\appendix
%\section{The ALICE Collaboration}
%\label{app:collab}
%\input{authors-list.tex}  %%%%%%% get the latest version before submitting

\begingroup
\small
\begin{flushleft}
J.~Adam\Irefn{org39}\And
D.~Adamov\'{a}\Irefn{org82}\And
M.M.~Aggarwal\Irefn{org86}\And
G.~Aglieri Rinella\Irefn{org36}\And
M.~Agnello\Irefn{org110}\And
N.~Agrawal\Irefn{org47}\And
Z.~Ahammed\Irefn{org130}\And
S.U.~Ahn\Irefn{org67}\And
I.~Aimo\Irefn{org93}\textsuperscript{,}\Irefn{org110}\And
S.~Aiola\Irefn{org135}
% ...
\renewcommand\labelenumi{\textsuperscript{\theenumi}~}

\section*{Affiliation notes}
\renewcommand\theenumi{\roman{enumi}}
\begin{Authlist}
\item \Adef{0}Deceased
\item \Adef{idp5842272}{Also at: University of Kansas, Lawrence, Kansas, United States}
\end{Authlist}

\section*{Collaboration Institutes}
\renewcommand\theenumi{\arabic{enumi}~}
\begin{Authlist}

\item \Idef{org1}A.I. Alikhanyan National Science Laboratory (Yerevan Physics Institute) Foundation, Yerevan, Armenia
\item \Idef{org2}Benem\'{e}rita Universidad Aut\'{o}noma de Puebla, Puebla, Mexico
\item \Idef{org3}Bogolyubov Institute for Theoretical Physics, Kiev, Ukraine
\item \Idef{org4}Bose Institute, Department of Physics and Centre for Astroparticle Physics and Space Science (CAPSS), Kolkata, India
\item \Idef{org5}Budker Institute for Nuclear Physics, Novosibirsk, Russia
\item \Idef{org6}California Polytechnic State University, San Luis Obispo, California, United States
\item \Idef{org7}Central China Normal University, Wuhan, China
\item \Idef{org8}Centre de Calcul de l'IN2P3, Villeurbanne, France
\item \Idef{org9}Centro de Aplicaciones Tecnol\'{o}gicas y Desarrollo Nuclear (CEADEN), Havana, Cuba
\item \Idef{org10}Centro de Investigaciones Energ\'{e}ticas Medioambientales y Tecnol\'{o}gicas (CIEMAT), Madrid, Spain
\item \Idef{org11}Centro de Investigaci\'{o}n y de Estudios Avanzados (CINVESTAV), Mexico City and M\'{e}rida, Mexico
\item \Idef{org12}Centro Fermi - Museo Storico della Fisica e Centro Studi e Ricerche ``Enrico Fermi'', Rome, Italy
\item \Idef{org13}Chicago State University, Chicago, Illinois, USA
\item \Idef{org14}China Institute of Atomic Energy, Beijing, China
\item \Idef{org15}Commissariat \`{a} l'Energie Atomique, IRFU, Saclay, France
\item \Idef{org16}COMSATS Institute of Information Technology (CIIT), Islamabad, Pakistan
\item \Idef{org17}Departamento de F\'{\i}sica de Part\'{\i}culas and IGFAE, Universidad de Santiago de Compostela, Santiago de Compostela, Spain
\item \Idef{org18}Department of Physics and Technology, University of Bergen, Bergen, Norway
\item \Idef{org19}Department of Physics, Aligarh Muslim University, Aligarh, India
\item \Idef{org20}Department of Physics, Ohio State University, Columbus, Ohio, United States
\item \Idef{org21}Department of Physics, Sejong University, Seoul, South Korea
\item \Idef{org22}Department of Physics, University of Oslo, Oslo, Norway
\item \Idef{org23}Dipartimento di Elettrotecnica ed Elettronica del Politecnico, Bari, Italy
\item \Idef{org24}Dipartimento di Fisica dell'Universit\`{a} 'La Sapienza' and Sezione INFN Rome, Italy
\item \Idef{org25}Dipartimento di Fisica dell'Universit\`{a} and Sezione INFN, Cagliari, Italy
\item \Idef{org26}Dipartimento di Fisica dell'Universit\`{a} and Sezione INFN, Trieste, Italy
\item \Idef{org27}Dipartimento di Fisica dell'Universit\`{a} and Sezione INFN, Turin, Italy
\item \Idef{org28}Dipartimento di Fisica e Astronomia dell'Universit\`{a} and Sezione INFN, Bologna, Italy
\item \Idef{org29}Dipartimento di Fisica e Astronomia dell'Universit\`{a} and Sezione INFN, Catania, Italy
\item \Idef{org30}Dipartimento di Fisica e Astronomia dell'Universit\`{a} and Sezione INFN, Padova, Italy
\item \Idef{org31}Dipartimento di Fisica `E.R.~Caianiello' dell'Universit\`{a} and Gruppo Collegato INFN, Salerno, Italy
\item \Idef{org32}Dipartimento di Scienze e Innovazione Tecnologica dell'Universit\`{a} del  Piemonte Orientale and Gruppo Collegato INFN, Alessandria, Italy
\item \Idef{org33}Dipartimento Interateneo di Fisica `M.~Merlin' and Sezione INFN, Bari, Italy
\item \Idef{org34}Division of Experimental High Energy Physics, University of Lund, Lund, Sweden
\item \Idef{org35}Eberhard Karls Universit\"{a}t T\"{u}bingen, T\"{u}bingen, Germany
\item \Idef{org36}European Organization for Nuclear Research (CERN), Geneva, Switzerland
\item \Idef{org37}Faculty of Engineering, Bergen University College, Bergen, Norway
\item \Idef{org38}Faculty of Mathematics, Physics and Informatics, Comenius University, Bratislava, Slovakia
\item \Idef{org39}Faculty of Nuclear Sciences and Physical Engineering, Czech Technical University in Prague, Prague, Czech Republic
\item \Idef{org40}Faculty of Science, P.J.~\v{S}af\'{a}rik University, Ko\v{s}ice, Slovakia
\item \Idef{org41}Faculty of Technology, Buskerud and Vestfold University College, Vestfold, Norway
\item \Idef{org42}Frankfurt Institute for Advanced Studies, Johann Wolfgang Goethe-Universit\"{a}t Frankfurt, Frankfurt, Germany
\item \Idef{org43}Gangneung-Wonju National University, Gangneung, South Korea
\item \Idef{org44}Gauhati University, Department of Physics, Guwahati, India
\item \Idef{org45}Helsinki Institute of Physics (HIP), Helsinki, Finland
\item \Idef{org46}Hiroshima University, Hiroshima, Japan
\item \Idef{org47}Indian Institute of Technology Bombay (IIT), Mumbai, India
\item \Idef{org48}Indian Institute of Technology Indore, Indore (IITI), India
\item \Idef{org49}Inha University, Incheon, South Korea
\item \Idef{org50}Institut de Physique Nucl\'eaire d'Orsay (IPNO), Universit\'e Paris-Sud, CNRS-IN2P3, Orsay, France
\item \Idef{org51}Institut f\"{u}r Informatik, Johann Wolfgang Goethe-Universit\"{a}t Frankfurt, Frankfurt, Germany
\item \Idef{org52}Institut f\"{u}r Kernphysik, Johann Wolfgang Goethe-Universit\"{a}t Frankfurt, Frankfurt, Germany
\item \Idef{org53}Institut f\"{u}r Kernphysik, Westf\"{a}lische Wilhelms-Universit\"{a}t M\"{u}nster, M\"{u}nster, Germany
\item \Idef{org54}Institut Pluridisciplinaire Hubert Curien (IPHC), Universit\'{e} de Strasbourg, CNRS-IN2P3, Strasbourg, France
\item \Idef{org55}Institute for Nuclear Research, Academy of Sciences, Moscow, Russia
\item \Idef{org56}Institute for Subatomic Physics of Utrecht University, Utrecht, Netherlands
\item \Idef{org57}Institute for Theoretical and Experimental Physics, Moscow, Russia
\item \Idef{org58}Institute of Experimental Physics, Slovak Academy of Sciences, Ko\v{s}ice, Slovakia
\item \Idef{org59}Institute of Physics, Academy of Sciences of the Czech Republic, Prague, Czech Republic
\item \Idef{org60}Institute of Physics, Bhubaneswar, India
\item \Idef{org61}Institute of Space Science (ISS), Bucharest, Romania
\item \Idef{org62}Instituto de Ciencias Nucleares, Universidad Nacional Aut\'{o}noma de M\'{e}xico, Mexico City, Mexico
\item \Idef{org63}Instituto de F\'{\i}sica, Universidad Nacional Aut\'{o}noma de M\'{e}xico, Mexico City, Mexico
\item \Idef{org64}iThemba LABS, National Research Foundation, Somerset West, South Africa
\item \Idef{org65}Joint Institute for Nuclear Research (JINR), Dubna, Russia
\item \Idef{org66}Konkuk University, Seoul, South Korea
\item \Idef{org67}Korea Institute of Science and Technology Information, Daejeon, South Korea
\item \Idef{org68}KTO Karatay University, Konya, Turkey
\item \Idef{org69}Laboratoire de Physique Corpusculaire (LPC), Clermont Universit\'{e}, Universit\'{e} Blaise Pascal, CNRS--IN2P3, Clermont-Ferrand, France
\item \Idef{org70}Laboratoire de Physique Subatomique et de Cosmologie, Universit\'{e} Grenoble-Alpes, CNRS-IN2P3, Grenoble, France
\item \Idef{org71}Laboratori Nazionali di Frascati, INFN, Frascati, Italy
\item \Idef{org72}Laboratori Nazionali di Legnaro, INFN, Legnaro, Italy
\item \Idef{org73}Lawrence Berkeley National Laboratory, Berkeley, California, United States
\item \Idef{org74}Lawrence Livermore National Laboratory, Livermore, California, United States
\item \Idef{org75}Moscow Engineering Physics Institute, Moscow, Russia
\item \Idef{org76}National Centre for Nuclear Studies, Warsaw, Poland
\item \Idef{org77}National Institute for Physics and Nuclear Engineering, Bucharest, Romania
\item \Idef{org78}National Institute of Science Education and Research, Bhubaneswar, India
\item \Idef{org79}Niels Bohr Institute, University of Copenhagen, Copenhagen, Denmark
\item \Idef{org80}Nikhef, National Institute for Subatomic Physics, Amsterdam, Netherlands
\item \Idef{org81}Nuclear Physics Group, STFC Daresbury Laboratory, Daresbury, United Kingdom
\item \Idef{org82}Nuclear Physics Institute, Academy of Sciences of the Czech Republic, \v{R}e\v{z} u Prahy, Czech Republic
\item \Idef{org83}Oak Ridge National Laboratory, Oak Ridge, Tennessee, United States
\item \Idef{org84}Petersburg Nuclear Physics Institute, Gatchina, Russia
\item \Idef{org85}Physics Department, Creighton University, Omaha, Nebraska, United States
\item \Idef{org86}Physics Department, Panjab University, Chandigarh, India
\item \Idef{org87}Physics Department, University of Athens, Athens, Greece
\item \Idef{org88}Physics Department, University of Cape Town, Cape Town, South Africa
\item \Idef{org89}Physics Department, University of Jammu, Jammu, India
\item \Idef{org90}Physics Department, University of Rajasthan, Jaipur, India
\item \Idef{org91}Physik Department, Technische Universit\"{a}t M\"{u}nchen, Munich, Germany
\item \Idef{org92}Physikalisches Institut, Ruprecht-Karls-Universit\"{a}t Heidelberg, Heidelberg, Germany
\item \Idef{org93}Politecnico di Torino, Turin, Italy
\item \Idef{org94}Purdue University, West Lafayette, Indiana, United States
\item \Idef{org95}Pusan National University, Pusan, South Korea
\item \Idef{org96}Research Division and ExtreMe Matter Institute EMMI, GSI Helmholtzzentrum f\"ur Schwerionenforschung, Darmstadt, Germany
\item \Idef{org97}Rudjer Bo\v{s}kovi\'{c} Institute, Zagreb, Croatia
\item \Idef{org98}Russian Federal Nuclear Center (VNIIEF), Sarov, Russia
\item \Idef{org99}Russian Research Centre Kurchatov Institute, Moscow, Russia
\item \Idef{org100}Saha Institute of Nuclear Physics, Kolkata, India
\item \Idef{org101}School of Physics and Astronomy, University of Birmingham, Birmingham, United Kingdom
\item \Idef{org102}Secci\'{o}n F\'{\i}sica, Departamento de Ciencias, Pontificia Universidad Cat\'{o}lica del Per\'{u}, Lima, Peru
\item \Idef{org103}Sezione INFN, Bari, Italy
\item \Idef{org104}Sezione INFN, Bologna, Italy
\item \Idef{org105}Sezione INFN, Cagliari, Italy
\item \Idef{org106}Sezione INFN, Catania, Italy
\item \Idef{org107}Sezione INFN, Padova, Italy
\item \Idef{org108}Sezione INFN, Rome, Italy
\item \Idef{org109}Sezione INFN, Trieste, Italy
\item \Idef{org110}Sezione INFN, Turin, Italy
\item \Idef{org111}SSC IHEP of NRC Kurchatov institute, Protvino, Russia
\item \Idef{org112}SUBATECH, Ecole des Mines de Nantes, Universit\'{e} de Nantes, CNRS-IN2P3, Nantes, France
\item \Idef{org113}Suranaree University of Technology, Nakhon Ratchasima, Thailand
\item \Idef{org114}Technical University of Split FESB, Split, Croatia
\item \Idef{org115}The Henryk Niewodniczanski Institute of Nuclear Physics, Polish Academy of Sciences, Cracow, Poland
\item \Idef{org116}The University of Texas at Austin, Physics Department, Austin, Texas, USA
\item \Idef{org117}Universidad Aut\'{o}noma de Sinaloa, Culiac\'{a}n, Mexico
\item \Idef{org118}Universidade de S\~{a}o Paulo (USP), S\~{a}o Paulo, Brazil
\item \Idef{org119}Universidade Estadual de Campinas (UNICAMP), Campinas, Brazil
\item \Idef{org120}University of Houston, Houston, Texas, United States
\item \Idef{org121}University of Jyv\"{a}skyl\"{a}, Jyv\"{a}skyl\"{a}, Finland
\item \Idef{org122}University of Liverpool, Liverpool, United Kingdom
\item \Idef{org123}University of Tennessee, Knoxville, Tennessee, United States
\item \Idef{org124}University of the Witwatersrand, Johannesburg, South Africa
\item \Idef{org125}University of Tokyo, Tokyo, Japan
\item \Idef{org126}University of Tsukuba, Tsukuba, Japan
\item \Idef{org127}University of Zagreb, Zagreb, Croatia
\item \Idef{org128}Universit\'{e} de Lyon, Universit\'{e} Lyon 1, CNRS/IN2P3, IPN-Lyon, Villeurbanne, France
\item \Idef{org129}V.~Fock Institute for Physics, St. Petersburg State University, St. Petersburg, Russia
\item \Idef{org130}Variable Energy Cyclotron Centre, Kolkata, India
\item \Idef{org131}Vin\v{c}a Institute of Nuclear Sciences, Belgrade, Serbia
\item \Idef{org132}Warsaw University of Technology, Warsaw, Poland
\item \Idef{org133}Wayne State University, Detroit, Michigan, United States
\item \Idef{org134}Wigner Research Centre for Physics, Hungarian Academy of Sciences, Budapest, Hungary
\item \Idef{org135}Yale University, New Haven, Connecticut, United States
\item \Idef{org136}Yonsei University, Seoul, South Korea
\item \Idef{org137}Zentrum f\"{u}r Technologietransfer und Telekommunikation (ZTT), Fachhochschule Worms, Worms, Germany
\end{Authlist}
\endgroup